\documentclass[11pt]{article}
\usepackage{graphicx}
\usepackage[margin=1.25in]{geometry}
\usepackage[usenames,dvipsnames]{color}
\usepackage{url}
\usepackage[colorlinks = true,
            linkcolor = blue,
            urlcolor  = blue,
            citecolor = blue,
            anchorcolor = blue]{hyperref}

\usepackage{enumitem}
\usepackage{lineno,hyperref,color}
\modulolinenumbers[5]
\usepackage{todonotes}
\usepackage{ulem}
\usepackage{booktabs}
\usepackage{float}
\usepackage{wrapfig}

\usepackage{appendix}
\usepackage{pdfpages}
\usepackage{hyperref}
\usepackage{geometry}                
\usepackage[parfill]{parskip}    
\usepackage{graphicx, subfigure}
\usepackage{amssymb}
\usepackage{amsmath}
\usepackage{epstopdf}
\usepackage{array}
\usepackage{lineno}
\usepackage{scrextend}
\usepackage{comment}

\def\Title#1{\begin{center} {\LARGE #1 } \end{center}}
\def\Author#1{\begin{center}{ \sc #1} \end{center}}

\begin{document}
\modulolinenumbers[1]

\Title{Comment for the SCAC Facilities Subcommittee from the Division of Particles and Fields of the American Physical Society}

\bigskip
\Author{
D.S. Akerib$^{1}$,
A. de Gouv\^ea$^{2}$,
P. Elmer$^{3}$,
S. Eno$^{4}$,
L. Fields$^{5}$,
A. Lankford$^{6}$,
V. Martinez Outschoorn$^{7}$,
P. Meade$^{8}$,
H. Murayama$^{9}$
}

\begin{center}
$^{1}$SLAC,
$^{2}$Northwestern University,
$^{3}$Princeton,
$^{4}$Chair of DPF, U. Maryland,
$^{5}$Notre Dame,
$^{6}$UC Irvine,
$^{7}$U. Mass. Amherst,
$^{8}$Stony Brook,
$^{9}$Chair Elect of DPF, UC Berkeley
\end{center}

\medskip

\def\thefootnote{\fnsymbol{footnote}}
\setcounter{footnote}{0}

\section{Introduction}

The 
Division of Particles and Fields (DPF) of the American Physical Society (APS)  would like to thank SCAC for providing an opportunity for input to the SCAC facilities subpanel.
The DPF  represents scientists working at U.S.\ Universities and  DOE Laboratories  on understanding fundamental particles and forces.  We are focused on Discovery Physics.
The subpanel has been asked to consider two proposed DPF-related projects, the DUNE Phase II Detectors and supporting infrastructure (\$1000M) and U.S.\ contributions to a future Higgs Factory, -- labeled as U.S. Contribution to Future Energy Frontier Collider [HEP] in the SCAC facilities summary document~\cite{SCAC} (\$1000M). The cost estimates are those provided to your committee via Ref.~\cite{SCAC}.

In this note,  we briefly describe the questions driving our field and the ``P5" strategic plan and other community prioritization documents that form the backbone of our comment.      The following sections provide short descriptions of the two projects, highlighting information relevant to your charge, answers to your specific questions, and a description of  the process that produced this note.

\section{Particle physics}

Particle Physics is about answering fundamental questions: what is everything made of, what is the nature of space and time, how did the universe get to look the way it does and where is it going? To answer these questions we use, adapt, and create new tools: accelerators, detectors, electronics, computing, data sharing and storage, data analysis methods, physics tools (quantum mechanics, quantum field theory), mathematics tools (string theory), {\it etc.} In the last 100+ years, societal investments in this area have paid off, creating both new knowledge and technical spin-offs. We expect that future investments will continue to do so. 
In addition, we attract and train brilliant people, most of whom use and adapt the skills they acquire -- outside of particle physics -- to improve society as a whole.

\section{Strategic planning for U.S.\ particle physics}\label{sec:planning}

Development of a scientific vision and discovery potential for particle physics has been performed on a more-or-less decadal basis as a two-step process. The first  is a two-year-long community exercise called “Snowmass,” organized by the DPF in conjunction with the agencies,  which culminates in an extensive report~\cite{butler2023report2021uscommunity}. The second step is a subcommittee (``P5") of the (former) High Energy Physics Advisory Panel, charged by the DOE Office of Science and NSF MPS to develop a strategic plan, including prioritization of projects/facilities, within constrained budget scenarios. The most recent strategic plan is the 2023 P5 report~\cite{Pp5} “{\it Exploring the Quantum Universe – Pathways to Innovation and Discovery in Particle Physics.”} Over 3,200 U.S.\ scientists signed an expression of support for this plan.
In 2025, at the request of the funding agencies, the National Academies also produced a consensus study report 
exploring the long-term goals of the field,
without consideration of costs~\cite{nam}.

The 2023 P5 report laid out three broad science themes, each with two science drivers: 
\begin{enumerate}[noitemsep]
	\item Decipher the Quantum Realm
	\begin{enumerate}[noitemsep]
		\item Elucidate the Mysteries of Neutrinos
		\item Reveal the Secrets of the Higgs Boson
	\end{enumerate}
	\item Illuminate the Hidden Universe
	\begin{enumerate}[noitemsep]
		\item Determine the Nature of Dark Matter
		\item Understand What Drives Cosmic Evolution
	\end{enumerate}
	\item Explore New Paradigms in Physics
	\begin{enumerate}[noitemsep]
		\item Search for Direct Evidence of New Particles
		\item Pursue Quantum Imprints of New Phenomena
	\end{enumerate}
\end{enumerate}

The strategic plan provides for a science-driven, balanced program across the breadth of the field, consisting of a portfolio of large, medium, and small projects that will continuously produce important physics results. It considers the technology R\&D required to enable new discoveries, training the nation’s next generation of public- and private-sector scientific and technical workers.

The highest priority recommendation by P5 is support for ongoing projects.
The ongoing projects Proton Improvement Plan II (PIP-II) and Long Baseline Neutrino Facility/Deep Underground Neutrino Experiment (LBNF/DUNEn Phase I) are
in the list of facilities under consideration by the Subcommittee. They
address science drivers 1(a) and 3(a) and 3(b).

The 2023 P5 strategic plan recommended five prioritized major new projects.
The first, CMB-S4, could not be realized as planned due to unexpected infrastructure constraints at the South Pole that were brought to the field's attention by NSF shortly after the release of the P5 report.  However, to pursue the scientific goals, although at a different level than originally envisioned, an alternative plan is  being developed that  upgrades existing facilities.  Therefore this is outside the scope of the current Subcommittee. 

The next two priorities, DUNE Phase II and an offshore Higgs factory (FCC-ee), are under consideration by the Subcommittee. 
The fourth project, a Generation 3 Dark Matter direct detection experiment, was not included in the list of proposed facilities and is discussed further in Section~\ref{sec:gaps}.
The fifth one is for consideration by the National Science Foundation, not DOE.

\subsection{DUNE Phase II}

 Phase II will make DUNE the definitive long-baseline neutrino experiment of its kind. It delivers three core scientific capabilities: a 75\% increase in beam intensity, a 50\% boost in far-detector mass, and an improved near detector, all relative to DUNE Phase I. It will provide the precision (statistics and systematics) required to uniquely address fundamental questions in particle physics not otherwise within reach. DUNE Phase II is a natural and necessary step to fully realize LBNF/DUNE Phase I investments, which include facilities such caverns and beam infrastructure to host the full scope of DUNE.

DUNE probes neutrino oscillations, the phenomenon where the three known types of neutrinos transform into one another as they travel. Phase II will allow researchers to precisely measure this quantum mechanical behavior across multiple transformation cycles, comprehensively testing the neutrino standard model with best-in-class precision.  Crucially, Phase II provides the sensitivity required to precisely determine whether  Charge-Parity (CP) symmetry is violated among the leptons. If CP violation occurs, neutrinos and antineutrinos behave differently as they oscillate. Discovering this difference may help answer one of cosmology's biggest mysteries: why is there so much ordinary matter, and virtually no antimatter, in the universe today? 
Additionally, Phase II expands DUNE’s capabilities to test the neutrino Standard Model, as a way to search for new physics in neutrino oscillations, including: non-unitary mixing, non-standard interactions, violation of charge, parity, and time reversal symmetry (CPT), and the possible existence of additional neutrino species. DUNE's capability to observe tau neutrino interactions in the beam, unique among current and future long- baseline experiments, constitutes a valuable additional handle for this.

Phase II also expands DUNE's capability to look for new phenomena, including those that may be responsible for neutrino masses, both in the near and the far detectors. 
The high-intensity beam along with the state-of-the-art near detectors serve as a facility to search for new particles, potentially related to the dark matter puzzle, produced alongside neutrinos. Innovations in detector readouts and a more effective shielding system would improve the visible energy threshold and the energy resolution of Phase-II far detector modules compared to Phase I modules. These improvements would fundamentally expand the physics opportunities with MeV-scale astrophysical neutrino sources, such as the sun and core-collapse supernova explosions. 

Since the prioritization of DUNE Phase II by P5, significant progress towards realization of Phase II has been made through R\&D and prototyping of technology under consideration for the near and far detectors.  A third run of the ProtoDUNE detector testbed at CERN will test improved charge and photon readout technologies for the far detector.  Construction and installation of Phase II components could begin in 2031 in a technically-limited scenario.  

As in Phase I, where international partners provided over 50\% of detector components, Phase II expects major global contributions. Coordinated R\&D is underway in Europe and through several national programs worldwide, complementing U.S.-led efforts.

\subsection{FCC-ee}

The FCC-ee~\cite{update} is a proposed 91\,km circular collider at CERN, costing approximately 15 BCHF.
The 2026 update of the European Strategy for Particle Physics~\cite{CERN-ESU-2025-002}, endorsed by the CERN Council, selected it as CERN's recommended future collider. The proposed $\sim \$1000$M U.S.\  contribution is consistent with P5's call for funding commensurate with U.S.\ participation in the LHC and HL-LHC.

 FCC-ee is designed to study the Higgs boson.  
Is it composite, connected to new forces or symmetries, or influenced by extra dimensions? Does it interact with dark matter? How did the Higgs field freeze throughout the Universe when it was a trillionth of a second old, and was this connected to the matter-antimatter asymmetry? The Higgs enables atoms to exist if, as predicted by the Standard Model (SM), it is responsible for the generation of the electron mass.  Does it couple to electrons?  Are there new particles to be discovered in its decays?
These are examples of the unanswered questions about this newest of the fundamental particles.

Precise measurement of the Higgs coupling to   known particles
will provide answers to these questions.  
New physics can affect observables via small quantum imprints -- systematically calculable deviations caused by particles or forces beyond the collider’s direct energy reach -- which can be revealed by precision measurements. 
The existence of the Higgs boson and the top quark were both inferred indirectly, and their masses estimated to tens of percent, at center-of-mass energies below their own mass.

Moreover, the energy span of FCC-ee enables studies of many other particles with an unprecedented precision. FCC-ee will produce almost a  \textit{million} times more comparable quality Z bosons, carriers of the weak nuclear force, than previously studied, indirectly probing
energy scales an order of magnitude beyond today's reach.

With full approval targeted for 2028, 
FCC-ee construction would begin in the early 2030s, and operation in the second half of the 2040s. 
The intellectual framework -- a Technical Design Report -- will be finalized within several years of approval, making early involvement essential.

The successful U.S.-CERN partnership on  LHC provides a strong foundation for future collaboration. The U.S.\ has delivered vital intellectual and material contributions to the LHC accelerator and to its major experiments, based on technologies and instrumentation developed and constructed in the U.S. 
Five U.S.\ scientists have had terms leading the two main LHC experiments. CERN's in-kind contributions to LBNF/DUNE demonstrate reciprocal benefits for U.S.-hosted projects. Continuing this partnership could create further opportunities for DUNE Phase II and the Electron-Ion Collider, which shares significant technology with the FCC-ee.

\section{HEP facilities and the Genesis Mission}

Both DUNE Phase II and FCC-ee represent critical infrastructure investments that will directly enable and advance the Genesis Mission's ambitious goals for AI-powered discovery science.  HEP is already deeply engaged with Genesis with contributions to 20 funded awards spanning AI-accelerated physics analysis, autonomous experiment operations, AI-driven detector design, and AI-driven detector readout. HEP infrastructure depends on radiation-hard, high-density custom microelectronics, including microelectronics operating at cryogenic temperatures, directly aligned with the Administration's priorities to strengthen domestic chip-design capabilities. HEP also contributes directly to the national discovery platform Genesis is building, through the American Science Cloud and federated data platforms. HEP's role as a massive, continuous data generator will only grow: DUNE Phase II and FCC-ee will produce exabyte-scale datasets from physical domains currently unavailable to AI researchers, including rare neutrino interactions, precision Higgs measurements, and exotic particle signatures. Such datasets will provide essential data for foundation models and AI agents. The AI systems being developed and tested on current facilities for operations and data analysis, and specific AI-enhanced detector design efforts, will enable us to design and operate the first truly AI-native future colliders and experiments. Future HEP facilities may use AI to accelerate design choices using differentiable/surrogate simulations, and active learning to co-optimize detectors and beams. By embedding AI as a core capability from conception through operation, DUNE Phase II and FCC-ee will 
showcase how large-scale discovery science is conducted in the AI era.

\section{The Subcommittee's specific questions}

\subsection{As you consider the challenges and opportunities facing science broadly in the US, what advice do you have for us as we approach our charge? What should be our highest priority consideration? What would you do if you were in our shoes?}

We support the rubric presented at the July 2026 SCAC meeting. 
The rubric listed 7 criteria, and appeared to give them equal weights.  We suggest  that ``potential for/needed to maintain U.S.\ leadership", ``potential for/ to accelerate scientific discovery", and ``needed to maintain national or economic security" have higher weight than the remaining criteria.
Beyond the facilities subpanel, our field needs a way to adjust priorities when discoveries present new directions.

\subsection{What gaps do you see in the list of proposed facilities we have been tasked to prioritize?}\label{sec:gaps}

The facilities in Section~\ref{sec:planning} are strongly supported by the P5 strategic plan, yet the subset before the Subcommittee leaves gaps in science, U.S.\ leadership, technology, and workforce.  
Because the Subcommittee's charge concerns facilities costing ($\gtrsim \$100$M), we focus here on gaps at that scale. 
These gaps are primarily due to budget constraints, but the gaps may become even more substantial if new scientific results make these areas even more compelling.  We highlight two significant gaps in the proposed facilities based on the P5~\cite{Pp5} and National Academy~\cite{nam} recommendations, but additional compelling facilities that cover discovery physics not accessible with the current program that are not presented to the subcommittee can be been found in Ref.~\cite{HEPAP}.

Section~\ref{sec:planning} noted that 
a Generation 3 dark-matter direct-detection experiment (G3), 
a high priority of the P5 strategic plan, is not under consideration.
G3 would use a large, ultra-low-background underground detector to search for weakly interacting dark matter, directly addressing science driver 2(a), Illuminate the Hidden Universe - Determine the Nature of Dark Matter.
Some G3 concepts could also search for neutrinoless double-beta decay, highlighted by the 2023 NSAC Long Range Plan~\cite{NSAC} and March 2026 Summaries report~\cite{SCAC}. 

A further gap is the absence of test facilities for the accelerator technologies needed to probe shorter distances and the physics underlying electroweak symmetry breaking. Because heavy new physics generally decouples from low-energy observables, accelerator advances remain essential both for direct discovery and for enhancing the reach of precision measurements. 

P5 recommendation 4 calls for “an aggressive R\&D program that, while technologically challenging, could yield revolutionary accelerator designs that chart a realistic path to a 10 TeV pCM collider.” This includes vigorous R\&D “based on proton, muon, or possible wakefield technologies… with a goal of being ready to build major test facilities and demonstrator facilities within the next 10 years.”  This priority was echoed by the 2025 National Academies EPP report~\cite{nam}, whose \textit{first} recommendation called for a national muon-collider R\&D program leading to a demonstrator of the key technologies. Such a demonstrator would be substantial in scale.
However, intermediate muon facilities would provide value without a collider, supporting nuclear and neutrino physics and providing a possible path beyond LBNF/DUNE phases towards a neutrino factory.  Combined with other general accelerator development test facilities as identified in P5 they will develop technology to support future high energy physics and broad applications in fusion, materials, resource exploration, and national security, while developing U.S. leadership and the workforce for future facilities.

\section{About this note}

This note was initiated by the chair line of DPF.  A call for writing-committee member nominations was sent to the DPF community. The authors of this document were chosen from the nominations by the DPF chair line.    
The document was circulated to DOE, the community, and to APS management for comments and updated for submission.

\bibliographystyle{elsarticle-num}
\bibliography{theBIB}

@misc{Pp5, 
   url="https://www.usparticlephysics.org/2023-p5-report/",
  title="{Pathways to Innovation and Discovery in Particle Physics}", 
  author="{The P5 Panel}",
  year="2023"
  }

@misc{nam, 
   url="https://nap.nationalacademies.org/catalog/28839/elementary-particle-physics-the-higgs-and-beyond",
  title="{Elementary Particle Physics: The Higgs and Beyond}",
  author="{National Academies of Sciences, Engineering, and Medicine}",
  year="2025"
  }

@article{update,
      title         = "{Update of the European Strategy for Particle Physics:
                       Remit of the European Strategy Group}",
      year          = "2024",
      url           = "https://cds.cern.ch/record/2908924",
}

@misc{butler2023report2021uscommunity,
      title="{Report of the 2021 U.S. Community Study on the Future of Particle Physics (Snowmass 2021) Summary Chapter}", 
      author={Joel N. Butler and others},
      year={2023},
      eprint={2301.06581},
      archivePrefix={arXiv},
      primaryClass={hep-ex},
      url={https://arxiv.org/abs/2301.06581}, 
}

@techreport{CERN-ESU-2025-002,
      title         = "{The European Strategy for Particle Physics: 2026 Update -
                       Recommendations by the European Strategy Group}",
      reportNumber  = "CERN-ESU-2025-002",
      address       = "Geneva",
      year          = "2025",
      url           = "https://cds.cern.ch/record/2950671",
      doi           = "10.17181/CERN.423R.S20Z",
}

@article{SCAC,
      title         = "{SCAC Facilities Summary Document}",
      year          = "2026",
      url           = "https://science.osti.gov/-/media/About/pdf/scac/reports/SC-Facilities-master-3-27-26.pdf",
}

@article{HEPAP,
      title         = "{HEPAP Facilities Subpanel Report}",
      year          = "2024",
      url           = "https://science.osti.gov/-/media/hep/hepap/pdf/Reports/2024/HEPAP-Facilities-Subpanel-Report-final-v11.pdf",
}

@article{NSAC,
      title         = "{ THE 2023 LONG RANGE PLAN FOR NUCLEAR SCIENCE}",
      year          = "2023",
      url           = "https://nuclearsciencefuture.org/wp-content/uploads/2024/03/23-G06476-2024-LRP-8.5x11-pcg-v1.5-3.14.24.pdf",
}

\end{document}